\documentclass{svjour3}                     % onecolumn (standard format)
\smartqed  % flush right qed marks, e.g. at end of proof
\usepackage{graphicx}
\begin{document}

\title{Production of $\Xi^-$-hypernuclei via the ($K^-,K^+$) reaction
in a quark-meson coupling model%\thanks{Grants or other notes
}


\author{K.~Tsushima \and R.~Shyam \and A.~W.~Thomas %etc.
}


\institute{
K.~Tsushima \at
CSSM, School of Chemistry and Physics, University of Adelaide, SA 5005, Australia\\
%Tel.: +61-8-83130624\\
%Fax: +61-8-83133551\\
\email{kazuo.tsushima@gmail.com}           %  \\
\and
R.~Shyam \at
Saha Institute of Nuclear Physics, 1/AF Bidhan Nagar, Kolkata 700064, India
\and
A.~W.~Thomas \at
CSSM and ARC Center of Excellence for Particle Physics at the Tera-scale,
School of Chemistry and Physics, University of Adelaide, SA 5005, Australia
}

\date{Received: date / Accepted: date}

\maketitle

\begin{abstract}
We study the production of $\Xi^-$-hypernuclei,
$^{12}{\!\!\!_{\Xi^-}}$Be and
$^{28}{\!\!\!_{\Xi^-}}$Mg,
via the ($K^-,K^+$) reaction
within a covariant effective Lagrangian
model, employing the bound $\Xi^-$ and proton spinors calculated
by the latest quark-meson coupling model.
The present treatment
yields the $0^\circ$ differential cross sections for the formation of simple s-state $\Xi^-$
particle-hole states peak at a beam momentum around 1.0 GeV/c
with a value in excess of 1 $\mu$b.
\keywords{$\Xi^-$-hypernuclei \and quark-meson coupling model \and cross sections}
\end{abstract}

\section{Introduction}
\label{intro}

The $(K^-,K^+)$ reaction is one of
the most promising ways of studying the $S = -2$ systems such as
$\Xi^-$-hypernuclei and a $H$ dibaryon.
As for $\Xi$ hypernuclei, although there are some hints of their
existence from emulsion events, no $\Xi$ bound
state was unambiguously observed in the few experiments performed through
the ($K^-,K^+)$ reaction on a $^{12}$C target.
However, in the near future experiments will be performed at JPARC to
search for the $\Xi^-$-hypernuclei via the $(K^-,K^+)$ reaction
with the best energy resolution of a few MeV and with large
statistics.%~\cite{nag06}.

The elementary cross sections for
$p(K^-, K^+)\Xi^-$ were measured in the 1960s
and early 1970s using hydrogen
bubble chambers.
In a recent study~\cite{shy11}, this  reaction was
investigated within a single-channel effective Lagrangian model where
contributions were included from the $s$-channel (Fig. 1(a)) and
$u$-channel diagrams which have as intermediate states $\Lambda$ and
$\Sigma$ together with eight three- and four-star
resonances with masses up to 2.0 GeV [$\Lambda(1405)$, $\Lambda(1520)$,
$\Lambda(1670)$, $\Lambda(1810)$, $\Lambda(1890)$, $\Sigma(1385)$,
$\Sigma(1670)$ and $\Sigma(1750)$, represented by $\Lambda^*$
and $\Sigma^*$ in Fig.~1a].
\begin{figure}[t]
\begin{center}
  \includegraphics[width=0.75\textwidth]{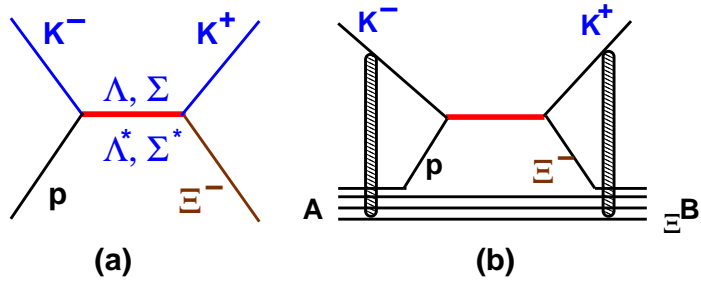}
\caption{
Graphical representation of the $p(K^-, K^+)\Xi^-$
(Fig.~1a) and $A(K^-, K^+)_{\Xi^-}B$ reactions (Fig. 2b) in our model.
In the latter case the shaded area depicts the optical model interactions in the
incoming and outgoing channels.}
\vspace{-2em}
\label{fig:1}       % Give a unique label
\end{center}
\end{figure}

Here we report on the $\Xi^-$-hypernuclear production via
$^{12}$C($K^-,K^+)^{12}{\!\!\!_{\Xi^-}}$Be  and
$^{28}$Si($K^-,K^+)^{28}{\!\!\!_{\Xi^-}}$Mg reactions~\cite{QMCXihyp},
within an effective Lagrangian model~\cite{shy04,shy08,shy09},
similar to that used in Ref.~\cite{shy11} to study the elementary
reaction, $p(K^-, K^+)\Xi^-$ combined with the quark-meson coupling (QMC)
model~\cite{gui88,gui08,tsu98b,sai96,sai07}.
We consider only the $s$-channel production diagrams
(Fig.~1b) as we are interested in the region $p_{K^-}<$ 2 GeV/c.
The bound $\Xi^-$ and proton spinors are calculated
in the latest version of QMC model~\cite{gui08}.
In this version, while the quality of results for $\Lambda$
and $\Xi$ is comparable to that of the earlier QMC results~\cite{tsu98b}, no
bound states for the $\Sigma$ states are found in middle and heavy
mass nuclei. The latter is in agreement with the experimental observations.
This is facilitated by the extra repulsion associated with the increased
one-gluon-exchange hyperfine interaction between the quarks in medium.

The use of the bound spinors obtained in the QMC model provides an
opportunity to investigate the role of the quark degrees of freedom in the
$\Xi^-$-hypernuclear production for the first time in studies
of this system. Since the $\Xi^-$-hypernuclear
production involves large momentum transfers (350 MeV/c - 600 MeV/c) to
the target nucleus, it is a good case for examining such short distance
effects. In the QMC model~\cite{gui88,gui08,tsu98b,sai96,sai07}, quarks within the
non-overlapping nucleon bags, interact
self-consistently with isoscalar-scalar ($\sigma$) and isoscalar-vector ($\omega$)
mesons in the mean field approximation. The explicit treatment of the
nucleon internal structure is a key in the model. The self-consistent response of the
bound quarks to the mean $\sigma$ field leads to a new saturation mechanism
for nuclear matter~\cite{gui88}. The QMC model has been used to study the
properties of finite nuclei~\cite{sai96}, the binding of $\omega$, $\eta$,
$\eta^\prime$ and $D$ nuclei~\cite{tsu98a,tsu98mesic,tsu99} and also the
effect of the medium on $K^\pm$ and $J/\Psi$ production~\cite{sai07}.

\section{Results and Discussions}

\begin{figure*}[t]
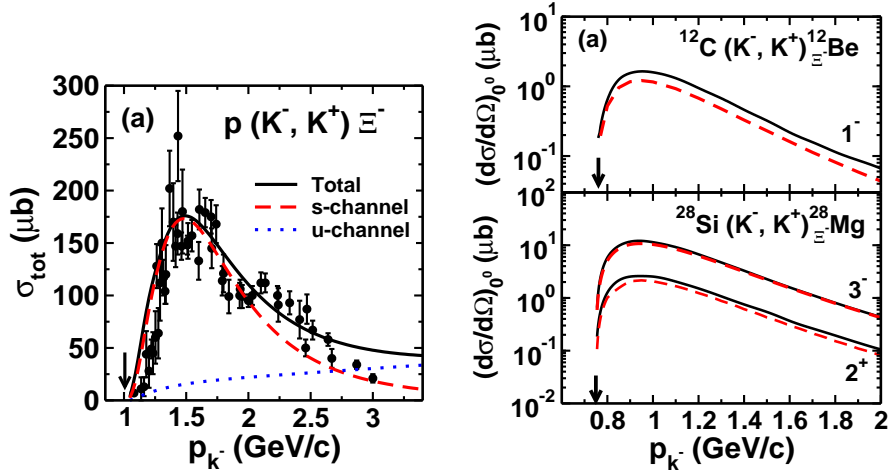

\begin{center}
\vspace{1em}
  \includegraphics[width=0.45\textwidth]{Fig3a.eps}\hspace{0.5cm}
  \includegraphics[width=0.45\textwidth]{Fig5a.eps}
\vspace{-3em}
\vspace{1cm}
\caption{
Total cross section for $p(K^-, K^+)\Xi^-$ (left panel in linear scale) and
the 0$^\circ$ cross sections for
$^{12}$C$(K^-,K^+)^{12}{\!\!\!_{\Xi^-}}$Be
and $^{28}$Si$(K^-,K^+)^{28}{\!\!\!_{\Xi^-}}$Mg (right panel in logarithmic scale, 
the solid [dashed] lines are the results of the QMC [phenomenological] model).
The arrows show the positions of the threshold beam momenta.
See Ref.~\cite{QMCXihyp} for details.
}
\vspace{-2em}
\label{fig:2}       % Give a unique label
\end{center}
\end{figure*}

In Fig.~2 (left panel), we compare our calculations with the
total cross section data of $p(K^-, K^+)\Xi^-$ for $K^-$ beam
momenta ($p_{K^-}$) below 3.5 GeV/c.  Our model can
describe well the beam momentum dependence of the elementary
total cross section data within statistical errors.
The measured total cross section peaks in
the region of 1.35-1.4 GeV/c which is well described by our model.

To calculate the cross sections for
$^{12}$C($K^-,K^+)^{12}{\!\!\!_{\Xi^-}}$Be  and
$^{28}$Si($K^-,K^+)^{28}{\!\!\!_{\Xi^-}}$Mg,
we have employed pure single-particle-single-hole $(\Xi p^{-1})$ wave
functions to describe the nuclear structure part~\cite{shy08},
ignoring any configuration mixing effects.
The amplitude involves the momentum space
Dirac-spinors of the bound $\Xi^-$ and proton.
We have used a plane wave approximation to
describe the relative motion of $K^-$ ($K^+$) in the incoming (outgoing) channel.
However, the distortion effects are partially accounted for by introducing
reduction factors to the cross sections as described in Ref.~\cite{ike94}.
%Our calculations are carried out all along in momentum space.

The thresholds for the $^{12}$C$(K^-,K^+)^{12}{\!\!\!_{\Xi^-}}$Be
and $^{28}$Si$(K^-,K^+)^{28}{\!\!\!_{\Xi^-}}$Mg reactions
are about 0.761 GeV/c and 0.750 GeV/c, respectively, and the
momentum transfers involved at 0$^\circ$, vary between 1.8 - 2.9
fm$^{-1}$. The initial states in both cases are doubly closed systems.
The QMC model predicts only one bound state for
the $^{12}{\!\!\!_{\Xi^-}}$Be system with the $\Xi^-$-1$s_{1/2}$ state
binding energy of 3.038 MeV.  For the $^{28}{\!\!\!_{\Xi^-}}$Mg
case, it predicts three distinct bound $\Xi^-$ states,
1$s_{1/2}$, 1$p_{3/2}$ and 1$p_{1/2}$, with the corresponding binding
energies 8.982, 4.079 and 4.414 MeV, respectively.

In case of the $^{12}$C target, $\Xi^-$-1$s_{1/2}$ state
can populate 1$^-$ and 2$^-$ states of the hypernucleus corresponding to
the particle-hole configuration $[(1p_{3/2})^{-1}_p,(1s_{1/2})_{\Xi^-}$].
The states populated for the $^{28}{\!\!\!_{\Xi^-}}$Mg hypernucleus are,
[2$^+$, 3$^+$], [1$^-$, 2$^-$, 3$^-$, 4$^-$], and [$2^-$, $3^-$]
corresponding to the configurations $[(1d_{5/2})^{-1}_p, (1s_{1/2})_{\Xi^-}$],
$[(1d_{5/2})^{-1}_p,(1p_{3/2})_{\Xi^-}$], and $[(1d_{5/2})^{-1}_p,
(1p_{1/2})_{\Xi^-}$], respectively.

In Fig.~2 (right panel), the 0$^\circ$ differential cross sections are shown as a function
of the beam momentum that are calculated by using the bound $\Xi^-$ and proton spinors
obtained in QMC (solid lines) as well as the phenomenological
model (dashed lines) for $^{12}$C$(K^-,K^+)^{12}{\!\!\!_{\Xi^-}}$Be
and $^{28}$Si$(K^-,K^+)^{28}{\!\!\!_{\Xi^-}}$Mg.
We have shown results for populating the
hypernuclear states with maximum spin and natural parity.
Although the relative motions
of $K^-$ and $K^+$ mesons respectively in the initial and final channels
are described by plane waves, the distortion effects
for the absorption of the incoming $K^-$ are included by
introducing factors that reduce the magnitudes of the cross sections. These
factors are taken to be 2.8 and 5.0 for $^{12}$C and $^{28}$Si targets,
respectively as suggested in Ref.~\cite{ike94}. This necessarily assumes
that shapes of the angular distributions are not affected by the distortion
effects. This aspect will be further investigated in a future study.

%\section{Conclusion}
%\label{conclusion}

Our results in Fig.~2 (right panel) show that for the both
reactions,
$^{12}$C($K^-,K^+)^{12}{\!\!\!_{\Xi^-}}$Be  and
$^{28}$Si($K^-,K^+)^{28}{\!\!\!_{\Xi^-}}$Mg,
the cross sections peak at $p_{K^-}$ around 1.0 GeV/c,
which is about 0.25-0.26 GeV/c above the corresponding production thresholds.  
This reflects the trends of the elementary $\Xi^-$ production reaction,   
where the peaks of the elementary total cross section as well as the zero 
degree differential cross section occur at about 0.35-0.40 GeV/c above 
the production threshold.
Furthermore, the magnitudes of the cross sections
near the peak position are in excess of 1 $\mu b$. It is important in this
context to note that the magnitude of our cross section for the $^{12}$C target
at a beam momentum of 1.6 GeV/c is similar to that obtained
in Ref.~\cite{ike94} within an impulse approximation model. Moreover, our cross
sections at 1.8 GeV/c also are very close to those of Ref.~\cite{gal83} for
the both targets. However, we fail to corroborate the results of
Ref.~\cite{gal83} where cross sections were shown to peak for $p_{K^-}$
around 1.8 GeV/c. It is quite probable that the distortion effects are
dependent on the beam momenta and may be relatively stronger at lower
values of $p_{K^-}$. Nevertheless, this is unlikely to lead to such a large
shift in the peak position. In any case, this effect was not considered in
Ref.~\cite{gal83} also. Thus, it seems necessary to re-examine the beam momentum
dependence of the zero-degree differential cross section in order to
understand this difference.

%\begin{table}
%\begin{tabular}{lll}
%\hline\noalign{\smallskip}
%\end{tabular}
%\end{table}

\begin{acknowledgements}
This work was supported by the University of Adelaide and the Australian
Research Council through grant FL0992247(AWT).
\end{acknowledgements}



\end{document}